\documentstyle[twoside,fleqn,espcrc2]{article}

\title{
\vspace{-10.0mm}
\begin{flushright}
\small
PC-561-09-97
\end{flushright}
        Finite size effects at phase transition in compact U(1) gauge theory
        \thanks{Contribution to Lattice '97, International Symposium, \
                Edinburgh, UK, 1997.}}

\author{Claude Roiesnel\address{Centre de Physique Th\'eorique,
        Ecole Polytechnique, \\
        Centre National de la Recherche Scientifique, UPR A0014, \\
        91128 Palaiseau Cedex, France}}
       
\begin{document}

\thispagestyle{empty}

\begin{abstract}
We present and discuss the results of a Monte-Carlo simulation of the
phase transition in pure compact U(1) lattice gauge theory with Wilson
action on a hypercubic lattice with periodic boundary conditions. The
statistics are large enough to make a thorough analysis of the size
dependence of the gap. In particular we find a non-zero latent heat in the
infinite volume limit.  We also find that the critical exponents $\nu$
and $\alpha$ are consistent with the hyperscaling relation but
confirm that the critical behavior is different from a conventional
first-order transition.
\end{abstract}

\maketitle

\section{INTRODUCTION}

The interest about the nature of the phase transition in compact 4D
U(1) lattice gauge theory has been revived by the recent development
of two new line of results. On the one hand, Kerler, Rebbi and Weber
\cite{KRW} studied the critical properties of the model by adding to
the standard Wilson action a coupling $\lambda$ controlling the
density of monopoles. They concluded to the existence of a
non-gaussian second-order critical point in the $(\beta, \lambda)$
plane. Damm and Kerler \cite{KD} are also investigating whether the
critical exponent of this transition is universal or changes with
$\lambda$.

On the other hand, Jers\`ak, Lang and Neuhaus \cite{JLN} studied the
compact U(1) gauge theory on lattices with sphere-like topology with a
Wilson action extended by a coupling $\gamma$ of charge 2
\begin{equation}
S=\beta \sum_{P} \cos \Theta_P + \gamma \sum_{P} \cos 2\Theta_P$$
\end{equation}
They found no gap on these lattices for $\gamma \leq 0$. They also
made a thorough finite-size size scaling analysis of their data and
concluded to the existence, for $\gamma \leq 0$, of a second-order
transition with a non-gaussian continuum limit.

Finally there is an investigation \cite{COX} of the scaling behaviour
of gauge-ball masses and of the static potential, which seems to
confirm the second-order nature of the transition also on lattices
with periodic boundary conditions at $\gamma = -0.2$ and $\gamma =
-0.5$.

However we must note that the critical exponents of these two
approaches are different. Moreover there is always an apparent
contradiction between the simulations on lattices with spherical
topology and on lattices with periodic boundary conditions since one
observes a gap on the latter even when $\gamma < 0$. One can fairly
state that some confusion about the nature of the transition still
persists.

Therefore it is useful to reconsider the simulation of compact pure 4D U(1)
lattice gauge theory with the standard Wilson action ($\gamma = 0$)
and periodic boundary conditions provided that such a simulation
fulfills two goals not met previously \cite{JNZ,FSS,GNC}:

\begin{itemize}

\item give an estimation of the infinite volume limit of the gap which is
      observed on finite-size lattices with periodic boundary conditions.

\item make a careful modern finite-size scaling analysis of the bulk
      critical behavior (as has been done on lattices with sphere-like
      topology \cite{JLN}).

\end{itemize}

In order to estimate the infinite volume limit of the gap, one needs
extrapolation formulas which require at least 3 parameters. Also the
asymptotic scaling formulas depend in general upon 3 parameters.
Determining these parameters from the measurement of one observable
only requires at least 6 to 7 data points. One needs more data points if
one wants to have some chance of estimating the subleading correction
terms. 

Therefore to be as systematic as possible the simulation has been done
on 9 lattice sizes from $L=4$ to $L=16$. One could argue that these
linear sizes are too small to reach the asymptotic scaling regime. If
this turns out to be indeed the case, then these smaller lattice sizes
are anyhow needed to determine the corrections to scaling which will be
required even when data on larger lattices become available.

\section{SIMULATION 1}

Since accurate data were lacking for many of the above lattice sizes,
a first simulation has been done to determine their pseudo-critical
coupling to about 1 part in $10^{-4}$. In order to reach this accuracy
a scanning of the pseudo-critical regions was done with a step
$\Delta\beta=10^{-4}$ and $10^{5}$ iterations at each coupling constant
$\beta$.

We used standard histogramming techniques to locate the double-peak
structure found in these simulations which is usually characteristic
of a first-order transition.  A pseudo-critical coupling was defined as
the coupling for which peaks have equal statistical weight. The latent
heat was defined as the gap between the two peaks at this
pseudo-critical coupling.

The details of this simulation are described in \cite{ROI} and we only give
here a short summary of the results.
 
A three-paramater fit to the gap $\Delta e(L)$ of the form
\begin{equation}
\label{fss}
\Delta e(L) = \Delta e(\infty) + a L^{-b} 
\end{equation}
gives $\Delta e(\infty) = 0.014(5)$ and $b = 1.03(16)$. We note that the
exponent $b$ is not consistent with the first-order prediction (at least in
Potts models) $b=D=4 \cite{BK,BGB}$. Fixing $b$ to 1 reproduces the data quite
well but with an infinite volume limit of the latent heat which is definitely
different from zero.  However it should be stressed that it is very difficult
to constrain the functional form of a three-parameter fit to the gap data.
Indeed an exponential fit
\begin{equation}
\label{exp}
\Delta e(L) = \Delta e(\infty) + a \exp(-b L) 
\end{equation}
reproduces the data as well with $\Delta e(\infty) = 0.0278(15)$.

An asymptotic finite-size scaling analysis of the pseudo-critical
couplings
\begin{equation}
  \label{bc}
  \beta_{c}(L)=\beta_{c}(\infty) + a L^{-\frac{1}{\nu}}
\end{equation}
yields the results $b_{c}(\infty)=1.01132(10)$ and $\nu=0.326(8)$.

We have also checked the scaling of the maxima of the specific heat
for lattice sizes in the range $L=4-12$. A two-parameter ansatz
\begin{equation}
C_{V,max}(L) = a L^{\frac{2}{\nu}-D}
\end{equation}
gives the independent determination $\nu=0.330(2)$ but with a rather high
$\chi^{2} \approx 3$ which hints at the need of correction terms to the
asymptotic formula.

All these results taken together confirm the rather paradoxical nature
of the U(1) phase transition. The critical exponents are completely
consistent with a second-order phase transition but with an index $\nu
\approx 0.33$ which is different from the value, $\nu \approx 0.36$,
quoted in \cite{JLN,COX}. This discrepancy raises the suspicion about
universality at different values of $\gamma$. On the other hand any
reasonable fitting ansatz to the gap data yields a non-zero value of
the latent heat in the infinite volume limit.  But again the approach
to this limit is different from the asymptotic formula expected within
the description of first-order transitions in the double gaussian
approximation \cite{BGB}. This disagreement might mean that the
asymptotic regime is not yet reached with lattice sizes up to $L=16$.

\section{SIMULATION 2}

Going to larger lattices is impracticable with local algorithms since already
we could not overcome the hysteresis on the $16^4$ lattice with $10^5$
iterations. However it is possible to attack the problem indirectly by
increasing the statistics on the smaller lattices so as to make a full finite
size analysis including corrections to scaling. The comparison of the finite
size scaling of several cumulants can unravel the systematic errors in the
critical exponents induced by the corrections to scaling. Combined fits can
reveal whether the critical exponents vary when excluding the smallest lattice
sizes.

Therefore we have made a second simulation at 3 to 5 coupling constants
selected in the pseudo-critical interval determined in simulation 1 at each
lattice size, except $L=16$. $10^6$ iterations have been done at each coupling
constant. These $10^6$ iterations were divided in two independent runs,
$5\times 10^5$ sweeps each, respectively from a hot start and a cold start,
using different random generators.

The data analysis, which is not yet completed, makes an extensive use of the
reweighting technique \cite{FS}. All independent runs on the same lattice size
are used as independent samples at the same $\beta$. The total amount of
statistics that we get is quite comparable to many of the Monte-Carlo
simulations of 3D spin models.

We are making a finite-size scaling analysis of 3 cumulants: the specific heat
per plaquette $C_{v}$, the Binder Cumulant $U_{4}$ and the second cumulant
$U_{2}$. We are also adding an analysis of their derivatives $dC_{v}/d\beta,
dU_{4}/d\beta, dU_{2}/d\beta$. These 6 cumulants are algebraically independent
and the position of their extrema defines a pseudo-critical coupling if
located in the scaling region. The value of each pseudo-critical coupling is
determined independently for every run by minimizing the corresponding
reweighted histogram with respect to $\beta$. Finally we take the statistical
average over all runs at each lattice size. The caveat of the method, and the
limiting factor of its applicability, is to ensure that the
Ferrenberg-Swendsen technique remains valid throughout the minimization
process.

\section{CONCLUSION} 

Preliminary results of the second simulation confirm the non-zero value of the
latent heat in the infinite volume limit. Fits for $\Delta e(\infty)$ with
Eq.~\ref{fss} and Eq.~\ref{exp} are completely consistent between both
simulations. However the value of the parameter $b$ in Eq.~\ref{fss} increases
to $\approx 1.40$ and the corresponding $\chi^2$ is much larger than for
the exponential fit which becomes highly favored.

Preliminary results from independent asymptotic finite size scaling fits to
the various definitions of the pseudo-critical couplings show deviations
$\approx 2\%$ among the values of the critical exponent $\nu$.  Even if these
deviations are much larger than the statistical errors, it will be very
difficult to extract the corrections to scaling.

\end{document}